\documentclass[transmag]{IEEEtran}

\usepackage{latexsym}
\usepackage{graphicx}
\usepackage{amsfonts,amssymb,amsmath}
\usepackage{tipa}
\usepackage{graphicx}
\usepackage{epsfig}
\usepackage{epstopdf}
\usepackage{upgreek}
\usepackage[nolist,nohyperlinks]{acronym}
\usepackage{multirow}
\usepackage{array}
\usepackage{subfigure}
\usepackage{cite}

\begin{acronym}
\acro{CNN}{Convolutional Neural Network}
\acro{NoC}{Network-on-Chip}
\acro{WNoC}{Wireless Network-on-Chip}
\acro{EM}{Electromagnetic}
\acro{SiO$_2$}{Silicon Dioxide}
\acro{AIN}{Aluminum nitride}
\acro{SiP}{System-in-Package}
\acro{SDM}{Software-defined metamaterial}
\acro{PDP}{Power Delay Profile}
\acro{mm-wave}{millimeter-wave}
\acro{MAC}{Medium Access Control}
\acro{OOK}{On-Off Keying}
\acro{ISI}{Inter-Symbol Interference}
\acro{TSV}{Through-Silicon Via}
\acro{CSMA}{Carrier Sense Multiple Access}
\acro{BER}{Bit Error Rate}
\acro{RZ}{Return-to-Zero}
\end{acronym}

\begin{document}

\title{Radiation pattern prediction for Metasurfaces: A Neural Network based approach}


\author{
Hamidreza Taghvaee, Akshay Jain, Xavier Timoneda, Christos Liaskos \\ Sergi Abadal, Eduard Alarc\'{o}n and  Albert Cabellos-Aparicio \\
\thanks{The authors gratefully acknowledge support from the EU's H2020 FET-OPEN program under grant No. 736876, and by the Catalan Institution for Research and Advanced Studies (ICREA).}
\thanks{Hamidreza Taghvaee, Akshay Jain, Xavier Timoneda, Sergi Abadal, Eduard Alarc\'{o}n and Albert Cabellos-Aparicio  are with NaNoNetworking Center in Catalunya (N3Cat) at Universitat Polit\`{e}cnica de Catalunya (UPC), Barcelona, Spain. Christos Liaskos is with Institute of Computer Science, FORTH, GR-71110 Heraklion, Crete, Greece. Akshay Jain is the corresponding author (Email: akshay.jain@upc.edu).}
}


\IEEEtitleabstractindextext{\begin{abstract}

As the current standardization for the 5G networks nears completion, work towards understanding the potential technologies for the 6G wireless networks is already underway. One of these potential technologies for the 6G networks are Reconfigurable Intelligent Surfaces (RISs). They offer unprecedented degrees of freedom towards engineering the wireless channel, i.e., the ability to modify the characteristics of the channel whenever and however required. Nevertheless, such properties demand that the response of the associated metasurface (MSF) is well understood under all possible operational conditions. While an understanding of the radiation pattern characteristics can be obtained through either analytical models or full wave simulations, they suffer from inaccuracy under certain conditions and extremely high computational complexity, respectively. Hence, in this paper we propose a novel neural networks based approach that enables a fast and accurate characterization of the MSF response. We analyze multiple scenarios and demonstrate the capabilities and utility of the proposed methodology. Concretely, we show that this method is able to learn and predict the parameters governing the reflected wave radiation pattern with an accuracy of a full wave simulation (98.8\%--99.8\%) and the time and computational complexity of an analytical model. The aforementioned result and methodology will be of specific importance for the design, fault tolerance and maintenance of the thousands of RISs that will be deployed in the 6G network environment. 

\end{abstract}


}

\maketitle
\IEEEdisplaynontitleabstractindextext


\acresetall


\section{Introduction}
Sixth-generation (6G) wireless networks will be even more heterogeneous and dense as compared to 5G and other legacy networks. Thus, the 6G architecture will need to be adapted to serve the ever evolving capacity and quality of service (QoS) requirements \cite{Zhang2019,Saad2019}. To satisfy these ever increasing demands, multiple enablers such as visible light communication (VLC), light fidelity (Li-Fi), Reconfigurable Intelligent Surfaces (RISs), TeraHertz (THz) communications, etc., have been proposed. Amongst these techniques, RISs have gained special attention. The reason being, through rapid tuning of the associated metasurfaces (MSFs), they transform the physical environment from being an adversary to being an ally in the communication process. Concretely, they enable more predictable and reliable propagation characteristics \cite{Basar2019,Renzo2019,DiRenzo2020,Papazafeiropoulos2020,Liaskos2018}. Such functionalities will be critical towards meeting the requirements being laid out for 6G networks \cite{Strinati2019, Elmeadawy2020,Rajatheva2020}.

The associated MSFs, in RISs, are electromagnetically thin-film and planar artificial structures, which have recently enabled the realization of novel electromagnetic (EM) and optical components with engineered and even atypical functionalities \cite{Li2020,DelHougne2020,Qian2020}. These include absorption of certain components of impinging radio signals (generated by the myriad transmitting devices/access points within the environment), as well as fine-grained manipulation of these radio signals in terms of direction, polarization, phase and power in a frequency-selective manner \cite{HABIB2006, Alitalo2009, Pendry2006, Shalaev2015}. On a more granular level, an MSF is composed of an array of subwavelength structures known as unit cells. Further, in this study we consider the case of tunable MSFs. From a general modeling perspective, unit cells in this case will consist of tunable resistors, \textit{R}, and capacitances, \textit{C}. This allows the unit cells to take multiple states and grants the MSFs their tunability characteristics. Notably, given a fixed target EM functionality, the design of an MSF is already a complex task. Hence, the design and operation of a tunable MSF will be even more challenging. A significant development in this regards has been made through multiple research efforts, such as \cite{Liaskos2018,Liu2018, Liaskos2018z}, which have discussed the possible architectures and characteristics of such programmable MSFs.

\begin{figure*}
    \centering
    \includegraphics[scale = 0.5]{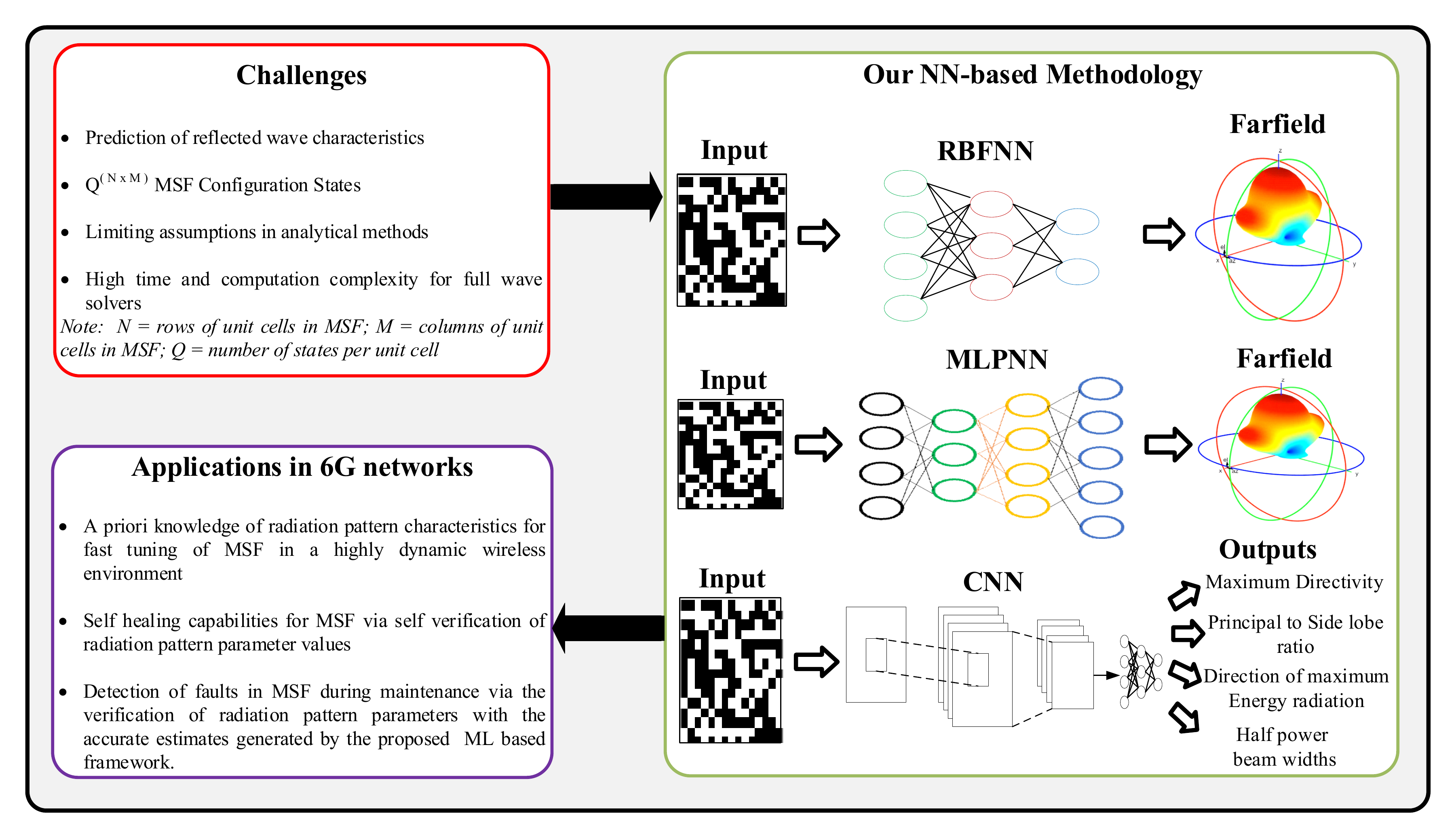}
    \caption{Research methodology outline system diagram.}
    \label{fig:Resout}
\end{figure*}


In addition, while tunability is definitely an advantageous property of the programmable MSFs, an important challenge associated with them is to be able to obtain the characteristics of a reflected wave given the parameters of the incident wave and the states of each composing unit cell, as shown in Fig. \ref{fig:Resout}. Moreover, as illustrated in Fig. \ref{fig:Resout}, a fast yet accurate estimation of the radiation pattern will facilitate multiple applications for 6G networks, such as the design, reliable functioning and maintenance of MSFs and consequently, RISs.
However, computing the characteristics of the reflected wave, given an MSF configuration, is presently challenging. The reason being that they are obtained by either utilizing analytical methods with multiple limiting assumptions or by conducting computationally intensive simulations through full wave EM solvers, as shown in Fig. \ref{fig:Resout}. 

To elaborate further, knowing the EM characteristics of each unit cell facilitates the calculation of the corresponding EM field. In most cases, the unit cell and thus the MSF is reflective (the transmission coefficient is zero). Further, we only need to have reflective features (reflection amplitude and phase) of the unit cell to obtain the Far Field pattern. Analytical models exist for describing and predicting the reflected EM field in some well-defined cases, such as beam steering and focusing of planar impinging waves. Still, these models introduce simplifications which can result into limited applicability in realistic setups and, consequently, reduced precision of results overall compared to the direct solution produced via Maxwell's equations~\cite{lagaris1998artificial}. Moreover, the iterative numerical full-wave simulations, which are widely adopted today and provide accurate device response predictions \cite{Hosseininejad2019a}, are severely memory-and time-consuming. Additionally, the design process largely relies on empirical reasoning or trial-and-error \cite{doi:10.1021/acs.nanolett.6b05137}, which is inefficient and often ineffective, especially when the problem is highly nonlinear. 

On the other hand, it is a well-known fact that machine learning (ML) techniques, and particularly Neural Networks (NNs), owing to their ability to learn complex relationships between input and output data are applicable of solving differential equations, thereby circumventing the need for numerical full-wave calculations~\cite{lagaris1998artificial,valasoulis2002solving,lagaris1998neural,tsoulos2009solving}. This fact provides the intuition towards another direction: since the MSF EM response (e.g., reflection) is essentially the solution to Maxwell's differential equations, it could be possible to design an ML construct that directly predicts the EM response, without resorting to full-wave simulations. 


Thus, this work provisions a data-driven NN based approach for determining an accurate estimation of the radiation pattern or several measures of interest that enable the full characterization of the radiation pattern. We now elaborate on the salient contributions of this paper, as follows:

\begin{itemize}
    \item We develop a novel Neural Network-based radiation pattern predictor, which, through our analysis, is established to be nearly as accurate as the full wave simulations but with the computational complexity of the analytical methods.
    \item To the best of our knowledge, this is a first method wherein certain important features of the reflected beam radiation pattern for a given MSF, i.e. \textit{Directivity, Principal-to-side-lobe ratio, Direction of maximum energy radiation} and \textit{Half power beam width}, have been predicted and effectively utilized for the complete characterization of the reflected beam radiation pattern. Consequently, this also provisions the applicability of our methodology in 6G networks (Fig. \ref{fig:Resout}).
    \item We provision a novel analysis based on the accuracy of prediction of the aforesaid parameters, for the locally tunable MSF scenario. Through the incremental design methodology, we establish a concrete framework and benchmark towards the selection of a CNN based predictor for the reflected beam radiation pattern. Specifically, we compare the performance of a CNN based predictor with an MLP based predictor. The comparative study reveals that the CNN predictor provisions an accuracy similar to the MLP predictor. It is imperative to state here that a CNN incurs significantly lower computational complexity as compared to an MLP neural network.  \\
\end{itemize}


\noindent The remainder of this paper is organized as follows: In Section II we present the current state of the art. In Section III we describe the incremental design framework, including the multiple scenarios that we have analyzed. In Section IV we elaborate upon the methodology that we have utilized for evaluating the multiple scenarios studied. In Section V we present the evaluation. We conclude the paper in Section VI.

\section{State of the Art}



\begin{figure}[t]
    \centering
    \includegraphics[scale = 0.47]{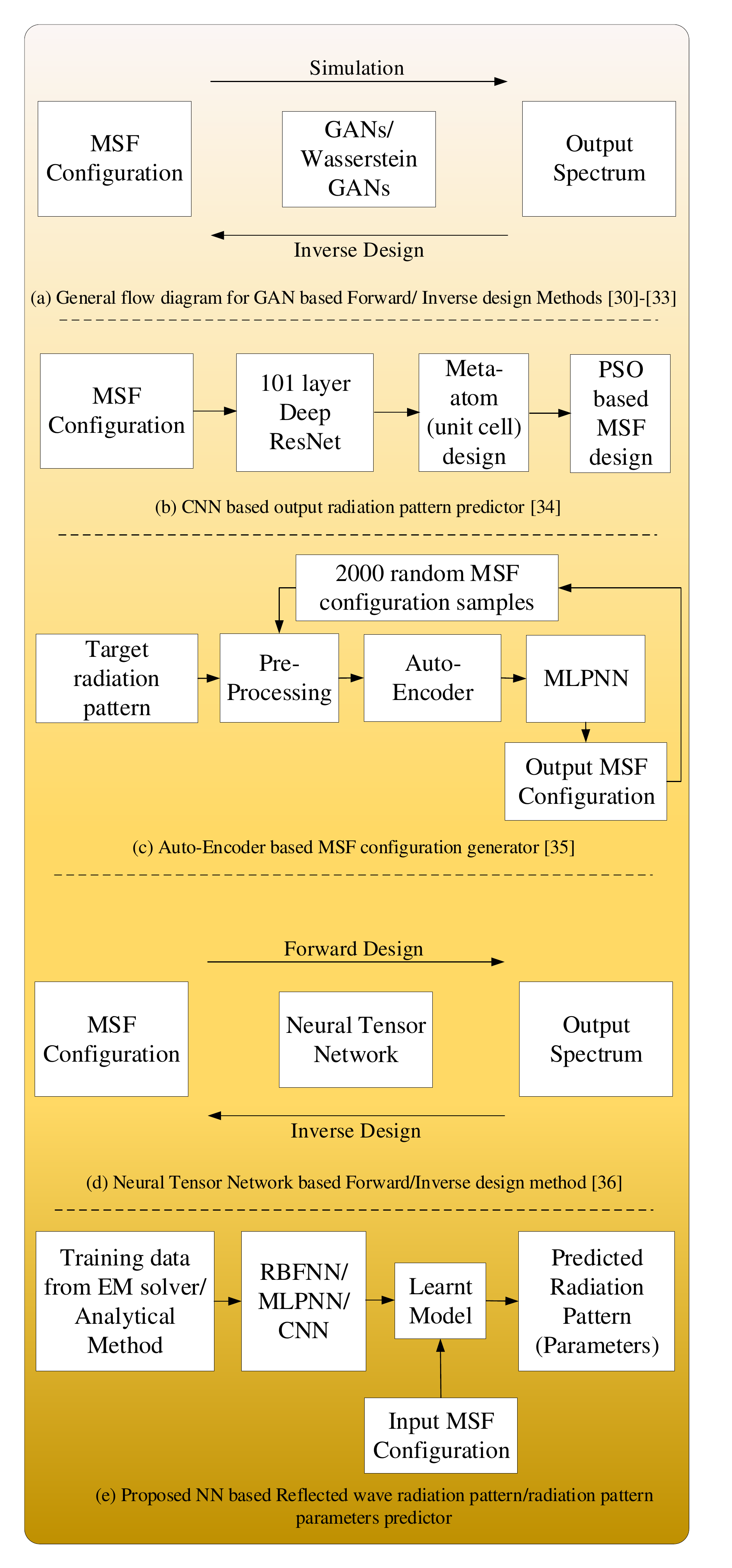}
    \caption{Schematic Representation of State of the Art approaches and the proposed method}
    \label{fig:Schema}
\end{figure}

ML methods over the past decade have gained significant importance in multiple sectors such as aerospace, medicine, or telecommunications \cite{Patil2017, Santos2016, Saad2003, Libbrecht2015, Mullainathan2017}. Further, since the laws of electromagnetism, fluid and aerodynamics are governed by well-known differential equation sets, the success of ML techniques in such domains was expectedly prospectful \cite{valasoulis2002solving, tsoulos2009solving, lagaris1998neural, lagaris1998artificial}.
 Specifically towards the design and validation of EM MSFs, which relate to the present paper, several works in the research community have recently proposed utilizing ML based algorithms for the same \cite{Liu2018b, Jiang2019, Hodge, An2019b,Zhang2019b, Qiu2019, An2019a}. We have consolidated these approaches into a schematic diagram and compared them with our proposed method in Fig. \ref{fig:Schema}. 
 
 In \cite{Liu2018b, Jiang2019, Hodge, An2019b} Generative Adversarial Networks (GANs) have been utilized to solve the inverse problem, i.e. to determine the MSF unit cell structure given a desired frequency response (Fig. \ref{fig:Schema}(a)). Additionally, in \cite{Liu2018b} a CNN is utilized as a simulator so as to be able to verify the accuracy of the frequency response of transmittance from the generated structures during the training phase of the GANs generator component. Similarly in \cite{Jiang2019}, a GAN based simulator, faster than the conventional numerical simulation tools, has been proposed. This simulator is one of the components of a system that does an inverse design to select a candidate unit cell metasurface pattern from a user-defined dataset of geometric structures, to match the required input optical spectrum. Additionally, in \cite{Hodge} GANs have been employed to design the metasurfaces that can generate complex tensorial RF responses. Further, and similar to previous methods, it also utilizes a CNN based simulator for the purposes of validating the RF response of the generated MSF configurations. Concretely, the CNN utilized simulates and generates the scattering paramaters for a given unit cell shape. However, the proposed simulator does not evaluate the complete radiation pattern of a locally or globally tunable MSF. Lastly, amongst the GAN based methods, \cite{An2019b} utilizes a variant of the conventional GANs, i.e. Wasserstein GAN (WGAN), to achieve its goal of identifying the most suitable MSF design. 

Next, in \cite{Zhang2019b}, an evolutionary algorithm that generates cell configurations and evaluates the fitness of each configuration by predicting the reflection phase with a trained CNN (a 101 layer Deep Residual network) for its given specific pattern has been proposed (Fig. \ref{fig:Schema}(b)). However, this CNN, which serves as a speedup of the optimization process of the evolutionary algorithm, is trained by previously encoding the output phases into a one-hot vector of length 360. Each element of the vector represents a discrete degree. Consequently, a problem that is purely based on regression is now converted into a classification problem. This, results in a loss of resolution and thus, crucial information with regards to the order and distance between degrees. Furthermore, in \cite{Zhang2019b}, the proposed CNN approach only provides good results for output radiation patterns with one, two or three beams. Therefore, for using this approach as a reflected phase predictor, the user needs to know \textit{a priori} how many lobes the resulting pattern will have. And given the fact that, a method to \textit{a priori} deduce the number of lobes has not been proposed in the aforementioned work, it thus limits the ability to generalize this approach to predict any reflected beam pattern. 


\begin{figure*}
    \centering
    \includegraphics[scale = 0.53]{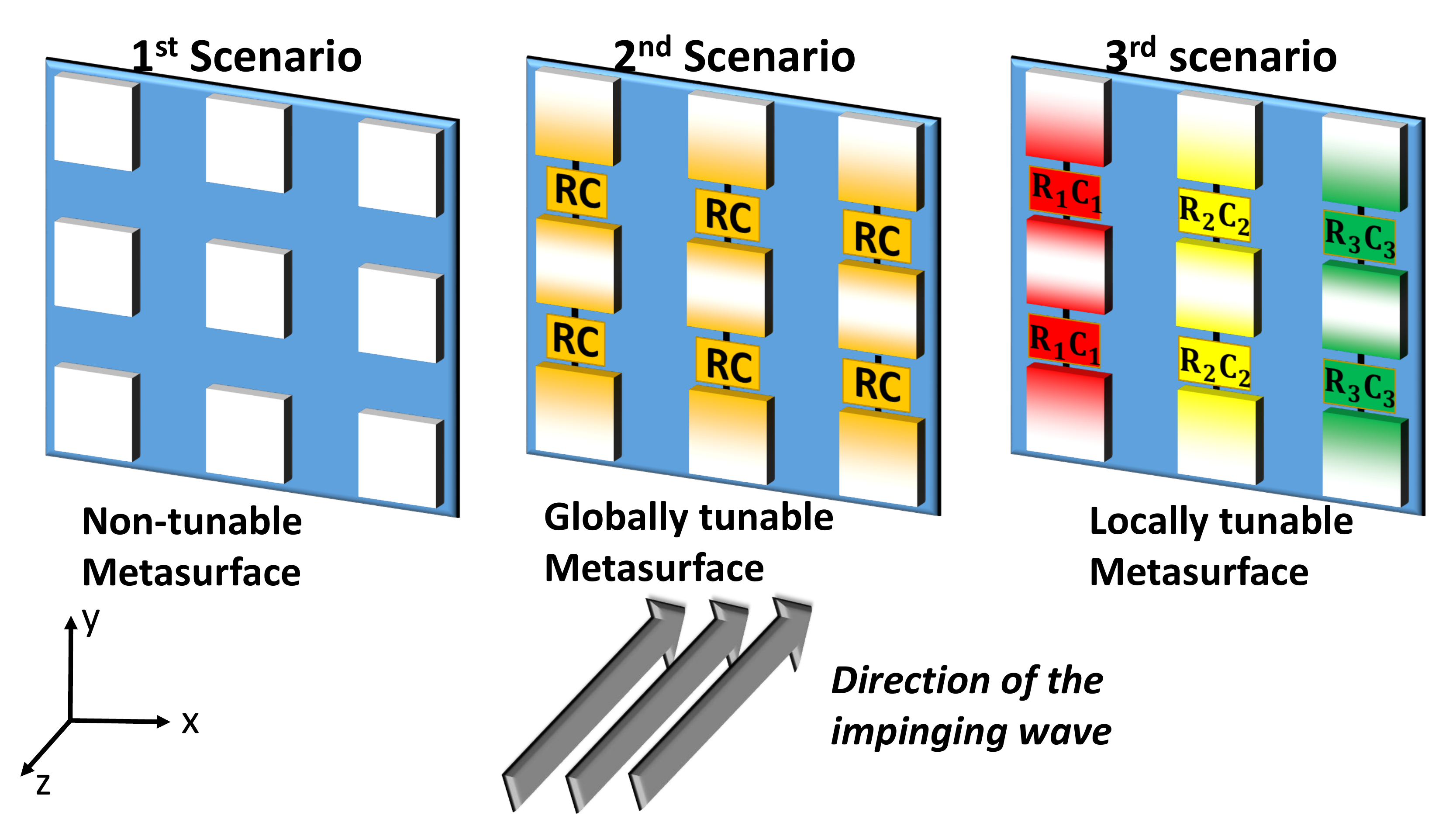}
    \caption{Diagram of the three scenarios utilized in the Incremental Design framework. First and Second scenarios correspond to the broader Homogeneous MSF configuration category, while the Third scenario corresponds to the Heterogeneous MSF configuration category.}
    \label{fig:MSscen}
\end{figure*}
Further, other research efforts such as \cite{Qiu2019} and \cite{An2019a} utilize other deep learning techniques to perform the task of MSF design. Concretely, in \cite{Qiu2019} an auto-encoder based approach has been adopted (Fig. \ref{fig:Schema}(c)). In this method, the auto-encoder enables capturing the most significant aspects of the input data, i.e., the desired reflected beam radiation spectrum, features. Subsequently, it facilitates the fully connected MLP network in determining the requisite metasurface structure for the demanded radiation pattern. Moreover, in \cite{An2019a}, a Neural Tensor Network (NTN) based approach has been adopted for designing the metasurface (Fig. \ref{fig:Schema}(d)). However, to do the same an initial simulation framework based on the NTN has been designed, which aims at predicting the amplitude and phase of the reflected wave from the MSF. This is performed by predicting the real and imaginary part of the desired EM response. Following this accurate prediction, inverse design methodologies are then adopted to formulate MSFs conforming to a wide variety of design objectives, thus highlighting the versatility of the proposed approach. Notably, other works, such as \cite{Wiecha2019},  have also used ML tools for solving different EM problems as a replacement of conventional numerical simulations. In \cite{Wiecha2019} an encoder-decoder structure was utilized for inferring the internal fields of arbitrary three-dimensional discretized nanostructures.    



And so, from Fig. \ref{fig:Schema}(e), it can be observed that our proposed methodology, which we will detail next, is unique compared to the state-of-the-art approaches (Figs. \ref{fig:Schema}(a)-(d)) in terms of its structure and approach towards predicting the output radiation pattern/radiation pattern parameters.

\section{Incremental Design Framework}\label{sec:scenarios}
We now describe the framework for our radiation pattern predictor, wherein we consider two broad scenarios, i.e., homogeneous and heterogeneous MSF configurations, and incrementally demonstrate that it is possible to predict the features of the reflected wave from a given MSF by means of data-driven learning approaches. Note that, depending on the scenario, the MSF is just a representation of a matrix of unit cells with given unit cell configurations. 

Furthermore, the homogeneous MSF configuration scenario is further expanded to two specific scenarios. These scenarios are established based on the underlying unit cell configurations of the MSF, and are listed as follows:

\begin{itemize}
    \item The first scenario consists of a non-tunable unit cell configuration across the MSF. Such a configuration is termed as a non-tunable MSF.  
    \item The second scenario consists of a matrix of unit cells across the MSF, wherein the unit cells have the same values for the tunable resistance \textit{R} and capacitance \textit{C}. Such a configuration is termed as a globally tunable MSF.
\end{itemize}

Subsequently, the heterogeneous MSF configuration scenario, or the locally tunable MSF, refers to the scenario where the unit cells can have different values of \textit{R} and \textit{C} associated with them. An illustration of these three scenarios that we have analyzed in this work, is presented in Fig. \ref{fig:MSscen}. We now describe these scenarios and the associated methodologies for radiation pattern prediction in detail through Sections III.A-B and IV, respectively. 



\subsection{Homogeneous MSF Configuration}

We now elaborate upon the two scenarios, i.e., the non-tunable MSF and the globally tunable MSF, in the text that follows.   
\subsubsection{First scenario (non-tunable, single unit cell / full radiation pattern estimation)} In the first scenario, we analyze whether the data-driven models are able to predict the complete reflected wave radiation pattern for a non-tunable single unit cell on a MSF. Note that, the MSF considered here consists of an infinite array of same unit cell configuration (Fig. \ref{fig:MSscen}, Case 1). Given the fact that the MSF is large enough ($>>\lambda/2$), it allows us to use periodic boundary conditions in the simulations, thus reducing the complexity. Further, in the prediction process, it is presumed that the azimuth and elevation angles of the incident EM wave are given. 
\subsubsection{Second scenario (tunable single unit cell / full radiation pattern estimation)}
    In the second scenario, we analyze whether the data-driven models are able to predict the complete reflected wave radiation pattern for a globally tunable MSF. One of the reasons for studying the globally tunable MSF configurations is the role that they will play in applications such as object tracking, sensing, radiation absorption, etc. And so, in this scenario the MSF consists of an infinite array of unit cells, wherein the same tuned unit cell configuration is repeated \textit{ad infinitum} (Fig. \ref{fig:MSscen}, Case 2).

\subsection{Heterogeneous MSF Configuration}
\subsubsection{Third scenario (tunable full surface / radiation pattern attribute estimation)} We now elaborate upon the third scenario, which expands our incremental framework to a locally tunable MSF (Fig. \ref{fig:MSscen}, Case 3). Such MSFs enable applications such as beam steering, beam focusing, etc., and hence, will be of significant importance in 6G networks. Thus, this reinforces our objective of studying and evaluating such MSF configurations. 

Concretely, and differing from the first and second scenarios, the MSF under study is composed of an array of unit cells that can have different states. Additionally, and again different from the first two scenarios, we evaluate whether our data-driven models are able to predict four measures of interest that characterize the complete reflected wave radiation pattern instead of the radiation pattern itself. These measures of interest are the \textit{Directivity}, \textit{Principal-to-side-lobe ratio}, \textit{Angle of maximum radiation} and \textit{Half power beam width}. Note that, evaluating only the aforementioned parameters helps to avoid having models that result in an output with very high dimensionality.
    
Next, in this scenario, the inputs for our NN based framework are two-dimensional matrices, with each value representing the 8 possible states of the unit cell at the corresponding position in the MSF. Additionally, the corresponding MSF is a $12\times12$ matrix of unit cells. The framework thus attempts to predict, for normal incident angles, the measures of the reflected beam radiation pattern for an MSF with a set of given unit cell state configuration. Note that, the number of possible configurations for the MSF under study, are $8^{144}$.



\section{Methodology}
\subsection{Homogeneous MSF Configuration}

\subsubsection{First Scenario}

For the simulations in the first scenario, we sweep the azimuth and elevation angles from 0 to 89 degrees with respect to normal incidence direction alongside a resolution of 1 degree. The reason for choosing this range of angles is that, given that the transmittance is 0, we do not need to evaluate negative elevation angles. Additionally, due to the assumed unit cell symmetries, we also do not need to explore all the azimuth angles.   

Further, the NN model that we explore for our data-driven framework was the Radial Basis Function Neural Network (RBFNN). As the name suggests, in an RBFNN the basis functions, which multiply the weights to determine the output of a layer in a neural network, are Gaussian. Therefore, the output of a hidden neuron is determined by the distance between the input and neuron’s center. Such a paradigm is \textit{a priori} very interesting for our approach, since it models spatial variables. This is in contrast to the MLPNN, in which the basis functions are based on the dot product. Concretely, this enables the RBFNN to learn the non-linear relationship between the incidence and reflection angles of the EM wave more effectively than an MLPNN, which is inherently based on an linear transformation. 

Note that, for the accuracy of evaluation of the RBFNN, we set the MSE goal to be $10^{-11}$, as stated above, and the spread constant to 1. Further, 8100 samples were collected using an EM simulator, of which 85\% were utilized for training and validation and the rest, i.e., 15\%, for evaluating the model generalization (which is usually referred to as the testing process). It is imperative to state here that, for non-deep learning scenarios, the aforementioned set of hyper-parameters lie within the range of values that are chosen usually \cite{Zhang2019b}.

\subsubsection{Second Scenario} 

In the second scenario, we vary the values of the parameters that characterize the physical structure of the unit cell, i.e., resistance \textit{R} and capacitance \textit{C}. However, as described earlier, the entire MSF consists of the same unit cell configuration throughout, i.e., all tuned unit cells will have the same value for \textit{R} and \textit{C}. Note that, for the sake of brevity, in the evaluations we assume that the incident wave direction to be normal. However, if required the evaluation of our model can be extended to any incident wave direction (incident angle). Further, we sweep the values of \textit{R} from $1 \Omega$ to $100 \Omega$ with a resolution of $1 \Omega$, and that of \textit{C} from 0.1 pF to 1 pF with a resolution of 0.01 pF. In addition, and importantly, the framework that we utilize in this scenario for our data driven approach is the MLPNN.

Moreover, unlike scenario 1, wherein the spatial characteristics of the incidence and reflected angles of the impinging wave was to be learnt, in scenario 2 the input features R and C lack any spatial characteristics. Thus, we do not evaluate RBFNN for this case. Furthermore, and owing to its relatively poor performance in scenario 1, we do not explore CNN for scenario 2. 

Next, for the MLPNN, we utilize a single hidden layer of 20 neurons. 
Further, the training algorithm used was Scaled Conjugate gradient without any regularization. The non-requirement of any regularization in our model was due to the fact that it has very small amount of parameters. In addition, and similar to the first scenario, we obtained the samples from an EM simulator and delimited 85\% of them for training and the rest for testing purposes. However, for this scenario we collected 9191 samples, which is slightly more than the number of samples collected for the first scenario.

\subsection{Heterogeneous MSF Configuration (Third Scenario)}
In this scenario, wherein we consider a locally tunable MSF, the samples we use for training and testing the model are collected through an analytical method, owing to the time and computational limitations. The reason being that, collecting enough number of samples through an EM simulator, so as to obtain a good model and given our computational power limitations, would take an extremely long period of time. Therefore, in this paper, and for this scenario, we demonstrate that: 

\begin{itemize}
    \item Our ML approach predicts the measures of the reflected beam pattern accurately.
    \item Provided that there is enough computational power, we can extrapolate the same model and methodology to the scenario where we have samples from an EM solver. \\
\end{itemize}
\begin{figure*}
    \centering
    \includegraphics[width=0.9\textwidth]{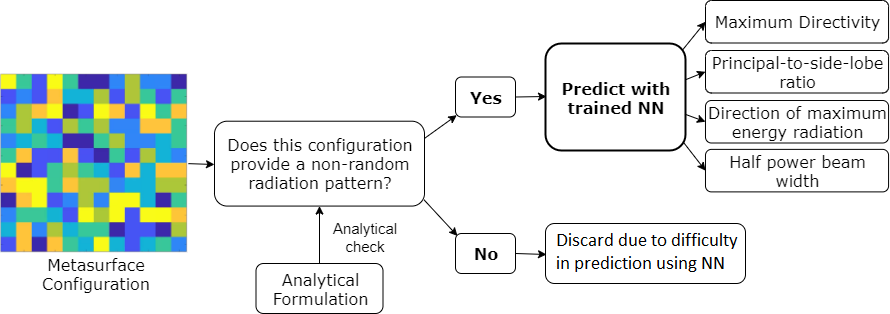}
    \caption{Diagram of the steps performed inside the system once the model is trained for the third scenario.}
    \label{fig:sysdiag}
\end{figure*}

\noindent \textbf{\textit{Analytical Model --}} The aforementioned analytical model computes the radiation pattern by applying the Huygens principle, wherein the far field is the sum of the contributions of all unit cells. Concretely, this model assumes that the crosstalk between adjacent unit cells can be neglected. We further assume that the MSF is uniformly illuminated by a normal incident plane wave and the reflection amplitude is constant across all the states. With these assumptions, we express the scattering field through eq. (1), as follows:
\begin{equation}
E(\theta, \phi) = K \sum_{i=1}^{M} \sum_{j=1}^{N}e^{j[\Phi_{ij}+k_0\zeta_{ij}(\theta, \phi)]},
\end{equation}
where $K$ is the reflection amplitude constant, $\Phi_{ij}$ is the reflection phase of unit cell $(i,j)$, $M$ and $N$ are the number of unit cells in a row or a column, $k_0$ is the wave number and $\zeta_{ij}(\theta, \phi)$ is the relative phase shift of the unit cells with respect to the radiation pattern coordinates ($\phi, \theta$) \cite{Hreza,Yang2016a}. Next, the relative phase shift $\zeta_{ij}(\theta, \phi)$ in eq. (1) is expressed as follows, 
\begin{equation}
\zeta_{ij}(\theta, \phi) = D_u\sin{\theta}[(i-\tfrac{1}{2})\cos{\phi}+(j-\tfrac{1}{2}) \sin{\phi}].
\end{equation}

This method has proven to be accurate in evaluating the far field of an MSF for beam steering by comparing the results with those of full-wave simulations \cite{Hosseininejad2019a}. And while, the approximations made have a small impact on the value and position of the side lobes, they are of minor relevance to the purpose of this work. \\



\noindent \textbf{\textit{Training and Testing dataset generation --}} Next, for ML, normally random selection is used to generate samples for training. However, random inputs of gradient for unit cell will always end up in a random scattering pattern. These patterns, in addition to being non-learnable, will not be of significance for design purposes. Thus, the samples collected for training are not entirely random combinations within the whole space, wherein the total number of combinations, as mentioned earlier, is $Q^{(N \times M)} = 8^{144}$.



Additionally, in our approach, a sample generation space is defined to control the entropy of the input data \cite{Cui2016}. 
Concretely, first a configuration that provides interpretative results, and hence without entropy, is randomly generated. Thereafter, entropy is introduced into the model with random ratio ranging from $[0-100]\%$. In this way, input data includes completely random samples as well as meaningful configurations. This vast range of entropy is precisely what is required to train our NN.


Subsequently, a criteria on the radiation pattern (e.g., Directivity) can be applied to discriminate interpretative configurations. This criteria translates the qualification of the NN on specific configuration. Therefore, we have a system that automatically checks if the new configurations used for predicting its measures can be used (with a reasonable granted accuracy) in the NN for prediction.

Utilizing the aforesaid process, the number of samples that were collected for training and testing the model of the third scenario was $10^5$. Amongst these samples, and similar to the first two scenarios, 85\% of them were used for training and validation whereas the remaining 15\% were stored in a completely separate set for the testing phase of the model. Further, from the training and validation set, 80\% of the samples were used for training, while the remaining 20\% were used for validation. The values of the pixels in the input images were normalized by performing a max-min escalation, without modifying their variance. This is not the case for our input variables, as the variance of each pixel is part of the relevant information the model uses for prediction.

However, it is important to state that standardizing the features is important when we compare measurements that have different units, as variables that are measured at different scales will not contribute equally and could end up creating a bias. And since this is the case for the target variables, the target samples for both training and testing sets were standardized by subtracting the mean of each of the measures and dividing them by their respective variances. \\

\noindent \textbf{\textit{Prediction System Operation --}} Following this, once our model is trained for a given upcoming configuration, it firstly estimates analytically whether the given configuration will provide interpretable results. If it does, it uses the trained model to predict the measures of interest. Instead, if the configuration outputs a random radiation pattern, it is discarded as the model cannot provide reliable results for this configuration. Fig. \ref{fig:sysdiag} illustrates the aforesaid steps performed in our system for predicting the measures of interest from an upcoming metasurface configuration, once the model is trained. \\

\noindent \textbf{\textit{NN Models --}} With this background, we now delve deeper into the setup of the two NN models that we utilize for our evaluations within the third scenario. 
\subsubsection{Multi-Layer Perceptron Neural Network}
As part of our methodology, illustrated in Fig. \ref{fig:sysdiag}, we utilize NNs for predicting the measures of interest of the reflected beam radiation pattern. Hence, in this section we consider MLP as our candidate NN. For the MLP case, the input images of $12\times12$ pixels which represent the unit cell configurations are flattened into vectors of 144 variables before being introduced into the NN.

\begin{figure} [!htb]
    \centering
    \includegraphics[width=0.48\textwidth]{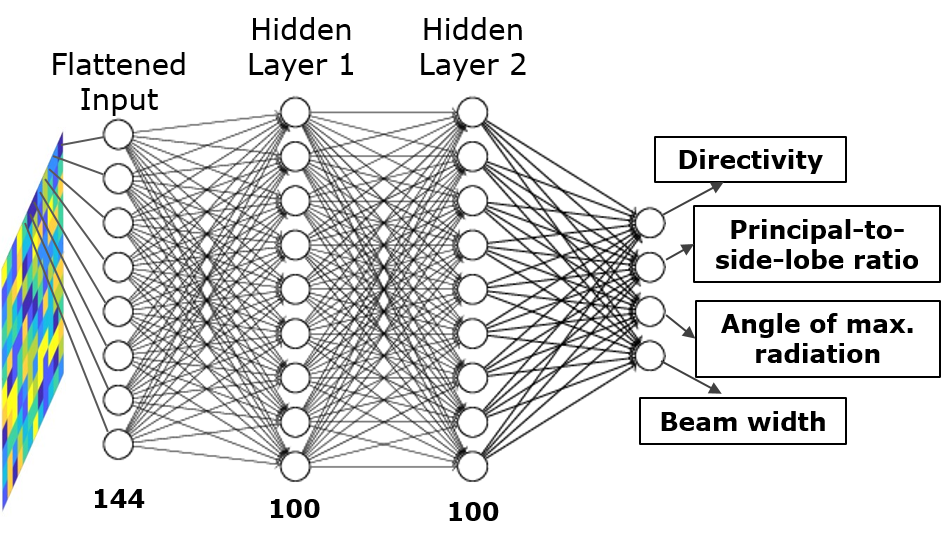}
    \caption{Structure of the Multi-Layer Perceptron Neural Network in the third scenario.}
    \label{fig:mlpstructure}
\end{figure}

\setlength{\parindent}{8pt}Fig. \ref{fig:mlpstructure} shows the structure of the MLPNN approach for the third scenario. The number of hidden layers and the neurons per layer were set to 2 and 100, respectively. A conclusion, with regards to the aforesaid parameter values, was reached after an extensive user-driven exploration, since sweeping across all the possible combinations was not computationally feasible. The rest of the parameters for the MLPNN are listed in Table I.  

\begin{table}[!htb]
\centering
\renewcommand{\arraystretch}{1.1}
\caption{Multi-Layer Perceptron Neural Network parameters}
\begin{tabular}{ |p{3.5cm}|p{3.5cm}| } 
 \hline
  \textbf{Parameter name} & \textbf{Value} \\
 \hline
  Regularization type & L2\\ \hline
  $\lambda$ & 0.8\\ \hline
  Training algorithm & Scaled Conjugate Gradient \\ \hline
  Number of hidden layers & 2\\ \hline
  Neurons of 1st hidden layer & 100\\ \hline
  Neurons of 2nd hidden layer & 100\\ \hline
 
\end{tabular}
\end{table}

As we can observe from Table I, the training algorithm selected is the Scaled Conjugate Gradient which accelerates the convergence rate with respect to first order algorithms, like the steepest descent, while avoiding the high computational cost of second order methods, such as the Newton's method. As the training time for our simplistic model is not a limitation, we can avoid the selection of the learning rate hyper-parameter, while we get better solutions in less iterations by using this quasi-Newton method. 

Next, regularization is a way to limit the complexity of a model and hence reduce the chances of overfitting by penalizing the most complex solutions in the cost function. Thus we employ an L2 regularization in our methodology. Specifically, in the model this is enforced via the $\lambda$ hyper-parameter. For a more detailed discussion on the regularization aspect, regularization method selection and the associated hyper-parameter value selection in our model, we refer the reader to Appendix A.

\subsubsection{Convolutional Neural Network}
Another NN that we explore for our methodology is the CNN. For the CNN case, the input images of $12\times12$ pixels and additionally a channel, which represents the unit cell configurations, are directly introduced to the NN.

\begin{figure}[!htb]
    \centering
    \includegraphics[width=0.45\textwidth]{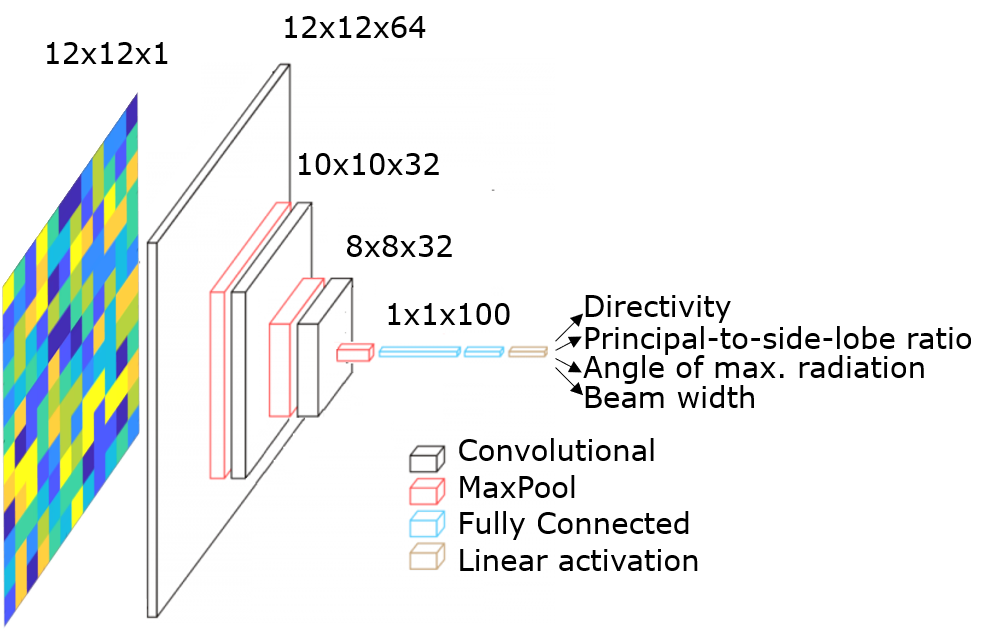}
    \caption{Structure of the Convolutional Neural Network in the third scenario.}
    \label{fig:cnnstructure}
\end{figure}

Fig. \ref{fig:cnnstructure} illustrates the structure of the CNN based approach for the third scenario. It is composed of three convolutional layers that consist of 64, 32 and 32 filters, respectively. Further, a max pooling process is performed after each of them. For all the convolutional layers, the filter size is $3\times3$ pixels and the stride is 1. As we do not use zero padding, the dimensionality of the intermediate images which represent the activations, is reduced at each layer. They are followed by a fully connected layer with 100 neurons, and an output layer with linear activation function. Similar to the MLPNN case, the architectural parameters of the CNN are a result of an extensive user-driven exploration, since sweeping around all the possible combinations was not computationally feasible. We enlist the most significant CNN architecture related parameters in Table II.  


\begin{table}[!htb]
\centering
\renewcommand{\arraystretch}{1.1}
\caption{Convolutional Neural Network architecture parameters}
\begin{tabular}{ |p{4cm}|p{3.5cm}| } 
 \hline
  \textbf{Parameter name} & \textbf{Value} \\
 \hline
  Regularization type & Dropout\\ \hline
  Dropout factor 3rd conv. layer & 0.2\\ \hline
  Dropout factor FC layer & 0.25\\ \hline
  Training algorithm & Stochastic Gradient Descent \\ \hline Learning rate & 0.001\\ \hline
  Momentum & 0.9\\ \hline
  Decay & 1e-4\\ \hline
  Num. of conv. layers & 3\\ \hline
  Num. of FC layers & 1\\ \hline

\end{tabular}
\end{table}
As we can observe, the training algorithm selected is the Stochastic Gradient descent (Table II). As it is a first order optimizer, the steps of the optimization process are linearly done with regards to the direction of maximum gradient. Thus, the length of the steps need to be defined by the learning rate hyper-parameter. The learning rate, decay, momentum, and the number of both convolutional and fully connected layers, specified in Table II, are set following a user-driven exploration. It is important to state here that, we do not use zero padding as it would introduce noise to the data, by essentially forcing a boundary that would be non-existent on a continuous metasurface plane.


Additionally, the third convolutional layer and the fully connected layer are regularized by means of a dropout process. This process consists of randomly ignoring a given number of layer outputs during the training process. Therefore, the layer with the dropout process is treated like a layer with lower number of nodes and connectivity to the previous layer. In effect, each update to a layer during training is performed with a different “view” of the configured layer. The parameter that controls the number of nodes which are randomly ignored is the dropout factor. 
For the third convolutional layer and the fully connected layer, the dropout factors are 0.2 and 0.25, respectively. These values were selected following the same procedure explained for selecting the $\lambda$ regularization parameter in the MLP.

\section{Evaluation} 
Given the framework discussed in Sections III and IV, we now present the evaluation for each of the scenarios discussed within this framework and highlight the relevant outcomes and insights. 

\subsection{Homogeneous MSF Configuration}
\subsubsection{First scenario}
\begin{figure}[!htb]
    \centering
    \includegraphics[scale = 0.27]{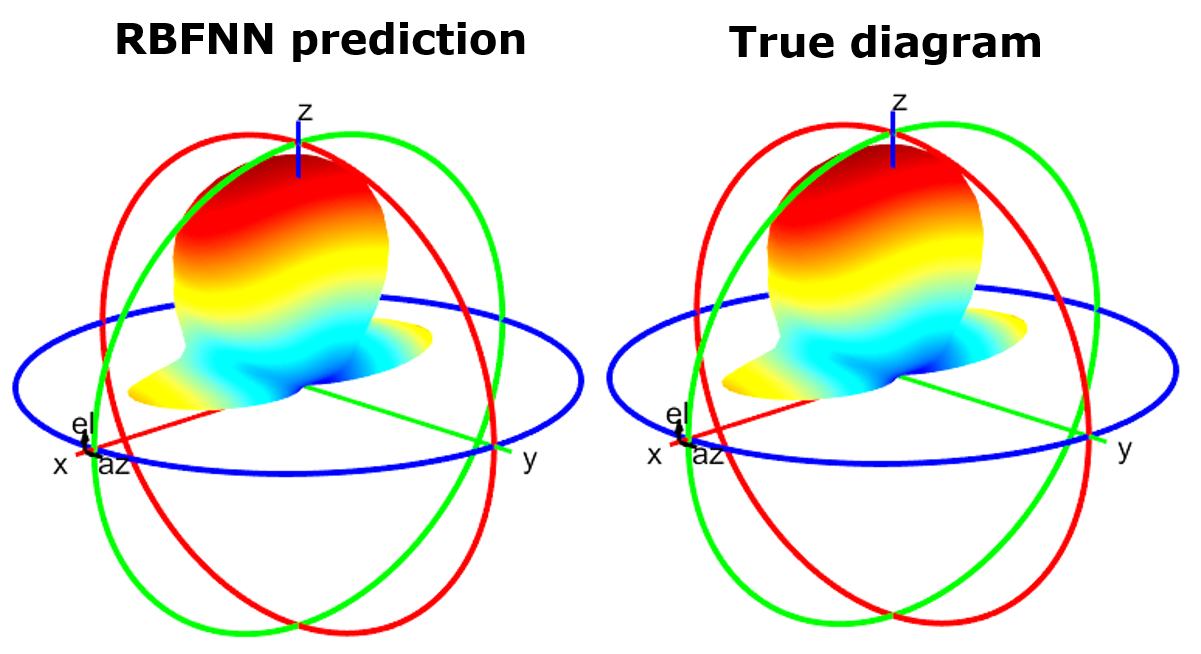}
    \caption{Comparison between the predicted radiation pattern by the RBFNN of the first scenario (left) and the true diagram (right) for azimuth an elevation values of 89.5 and 88.7 degrees with respect to the normal direction, respectively.}
    \label{fig:1stscenariores}
\end{figure}

For the non-tunable, single unit cell/ full radiation pattern case, the trained RBFNN was able to predict the radiation pattern for any given angle of incidence with an $R^{2}$ test of 0.9994. Therefore, this assists us in validating our hypothesis that ML models are able to accurately predict the reflected wave radiation pattern from a single unit cell for every angle of the incident wave. Fig. \ref{fig:1stscenariores} illustrates a visual comparison between the predicted radiation pattern by the trained RBFNN and the true diagram obtained through EM simulation, for the azimuth and elevation angles that were not present in the training set. This also emphasizes upon the fact that, our prediction system is able to accurately learn and generalize for untrained/unseen angles within the training dataset.    

Further, when a CNN was applied for this case, the observed mean squared error (MSE) was $10^{-7}$, which is significantly worse as compared to the accuracy obtained via the RBFNN approach (the MSE goal to measure the RBFNN accuracy was set to $10^{-11}$). Hence, for the sake of brevity, for scenario 1 we only highlight the results from the evaluations carried out using the RBFNN approach.   

\subsubsection{Second scenario}

\begin{figure}[!htb]
    \centering
    \includegraphics[scale = 0.27]{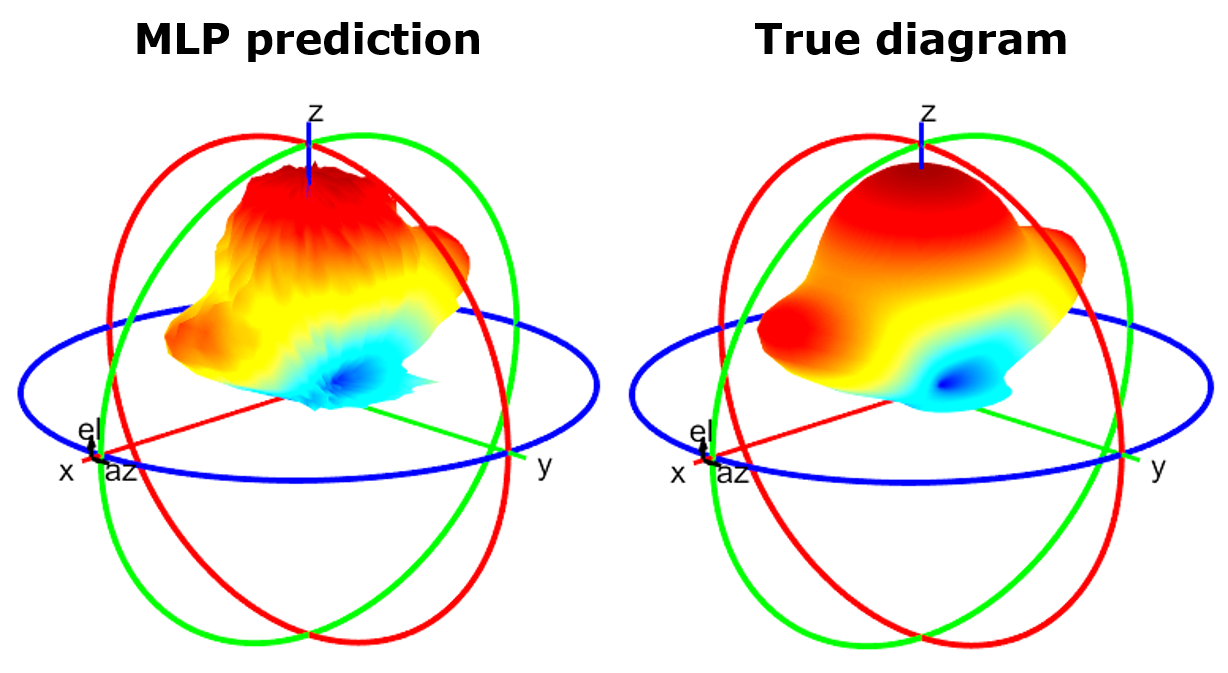}
    \caption{Comparison between the predicted radiation pattern by the MLP of the second scenario (left) and the true diagram (right) for R and C values of $2.5 \Omega$ and 0.25 pF, respectively.}
    \label{fig:2ndscenariores}
\end{figure}

For the tunable, single unit cell / full radiation pattern case, the trained MLP was able to predict the radiation pattern for any given R and C value with an $R^{2}$ test of 0.9849. Therefore, our hypothesis that, ML models can accurately predict the radiation pattern of the reflected wave in a single unit cell for each R and C combination, has also been validated. Fig. \ref{fig:2ndscenariores} shows the visual comparison between the predicted radiation pattern by the trained MLPNN and the true diagram obtained through EM simulation, for R and C values that were not present in the training set. This reinforces the fact that, our predictor is able to learn and generalize to scenarios with untrained/unseen values of R and C within the training dataset. 

Further, here we do not present a discussion of results for this scenario with the RBFNN and CNN setups. Specifically, given that RBFNN is not suitable for the second scenario (as discussed in Section IV.A.2), and the CNN performs extremely poorly for the first scenario, we do not detail a discussion on the performance of these setups here.   

\subsection{Heterogeneous MSF Configuration (Third scenario)}

The radiation pattern \emph{attribute} prediction problem for the third scenario, is essentially a regression problem. Hence, the cost/error function to minimize during the training process is the MSE. However, this error function does not provide very good interpretability of the performance. Alternatively, we define a tolerance (or a set of tolerances) specific for each measure of interest. Subsequently, we evaluate the percentage of the predictions that fall within this tolerance limit. This is also termed as the accuracy measure in this paper. Thus, in the following sections (V.B.1-V.B.4) we discuss the performance of the MLPNN and CNN over the different measures of interest that we aim to predict utilzing our methodology (Fig. 4). The results associated with the ensuing discussions have also been illustrated in Table III. 

\subsubsection{Directivity}
For the case of the \textit{Directivity} parameter, we observed that the MLPNN provided near perfect prediction, subject to certain tolerance limits. Concretely, from Table III, it can be seen that 95\% of the test samples have been accurately predicted when the tolerance is set to 0.25 dB. In addition, when the tolerance is relaxed further, i.e., to 0.5 dB, we observe an improved accuracy of 99.99\%. However, when the tolerance limit is reduced, i.e., to 0.1 dB, we notice that the accuracy of the MLPNN degrades drastically to 56.3\%.  

On the other hand, when we utilized the CNN as our predictor, the accuracy of prediction with a 0.25 dB tolerance limit was 90.6\% (Table III). Further, when we increased the tolerance to 0.5 dB, the accuracy of prediction improved to 99.8\%. Additionally, when we reduced the tolerance limit to 0.1 dB, similar to the MLPNN case, the accuracy of prediction for the CNN deteriorated significantly to 48.8\%. Note that, these aforementioned accuracy measures are lower than those offered by the MLPNN. This is because, an MLP based method, due to the fully connected architecture, can learn almost any feature space accurately. On the other hand, a CNN tries to extract the most significant features through its convolution based processing and hence, is a lossy method. 

However, a point of contention with the MLPNN is that, the fully connected architecture is not scalable for bigger MSF configurations. This will progressively become detrimental to the system performance, as the cost of computation will increase exponentially. In contrast, a CNN utilizes significantly less computational and memory resources and will scale better, whilst providing an accuracy measure that is close to that offered by the MLPNN. 
\begin{table}[t]
\caption{Accuracy Measure: MLPNN vs CNN}
\centering
\renewcommand{\arraystretch}{1.1}
\begin{tabular}{|>{\centering}p{2.5cm}|c|c||c|c|}
\hline
\multirow{2}{*}[0.1em]{\textbf{Parameter}} &  \multicolumn{2}{|c||}{\textbf{MLPNN}} & \multicolumn{2}{|c|}{\textbf{CNN}}  \\ \cline{2-5}
& Tolerance & Accuracy & Tolerance & Accuracy \\ \hline 
\multirow{3}{*}[0.1em]{Directivity}& 0.5 dB & 0.999 & 0.5dB & 0.998 \\ \cline{2-5}
& 0.25 dB & 0.950 & 0.25dB & 0.906 \\ \cline{2-5}
& 0.1 dB & 0.563 & 0.1dB & 0.488 \\ \hline
\multirow{3}{*}[0.7em]{Principle-to-side}& 0.5 dB & 0.999 & 0.5dB & 0.994 \\ \cline{2-5}
\multirow{1}{*}[-0.3em]{lobe ratio} & 0.25 dB & 0.983 & 0.25dB & 0.943 \\ \cline{2-5}
& 0.1 dB & 0.861 & 0.1dB & 0.801 \\ \hline
\multirow{3}{*}[0.7em]{Angle of maximum}& 5º & 0.998 & 5º & 0.989 \\ \cline{2-5}
\multirow{1}{*}[-0.3em]{radiation} & 2º & 0.727 & 2º & 0.607 \\ \cline{2-5}
& 1º & 0.406 & 1º & 0.319 \\ \hline
\multirow{3}{*}[0.1em]{Beam Width}& 1º & 0.995 & 1º & 0.988 \\ \cline{2-5}
& 0.5º & 0.973 & 0.5º & 0.926 \\ \cline{2-5}
& 0.25º & 0.792 & 0.25º & 0.618 \\ \hline
\end{tabular}
\end{table}

\subsubsection{Principle-to-side lobe ratio}

For the \textit{Principle-to-side lobe ratio}, we obtain similar observations from Table III, as we did for the \textit{Directivity} parameter. Specifically, for the MLPNN, when we vary the tolerance from 0.5 dB to 0.25 dB and finally to 0.1 dB, the corresponding accuracy measures are registered at 99.9\%, 98.3\% and 86.1\%, respectively. On the other hand, for the same tolerance value ensemble, the CNN method produces accuracy measures of 99.4\%, 94.3\% and 80.1\%, respectively. 

And so, as we can see that the MLPNN performs slightly better than the CNN. However, as mentioned earlier, this comes at a significant computational cost, thus hampering its scalability. In addition, 
it is understood that the correlation between the neighboring unit cells is far less as compared to those that are found in images in general \cite{Zhang2019b}. Consequently, this corroborates the findings from Table III, with regards to the CNN performing slightly worse as compared to the MLPNN. Concretely, an MLPNN can learn the interactions between the distinctly related neighboring unit cells much more effectively due to the fully connected architecture. However, a CNN treats the MSF like an image, thus considering the neighboring unit cells to be correlated. However, in reality this is seldom the case.

It is imperative to state here that, the aforesaid non-relational nature of nearby unit cells is also responsible for the visibly subdued performance of the CNN, as compared to the MLPNN, for other measures of interest.

\subsubsection{Angle of Maximum Radiation}
The results for the angle of maximum radiation in Table III are obtained by averaging the accuracy of prediction of the elevation and azimuth angles, with the purpose of providing a single view over this feature. Subsequently, we observe that the MLPNN performs slightly better than the CNN, the reasons for which have been expressed in Section V.B.2. 


To elaborate, for this measure we consider tolerance values of 5º, 2º and 1º. Next, from Table III we observe that the MLPNN has an accuracy of 99.8\%, 72.7\% and 40.6\% for the corresponding tolerance values, respectively. Further, the CNN approach has an accuracy of 98.9\%, 60.7\% and 31.9\%, given the same tolerance value ensemble, respectively. As can be seen, the accuracy drops as we reduce the tolerance limit, which is inline with our observations from the other measures of interest so far. Additionally, it can be deduced that irrespective of the NN utilized for the prediction step, the accuracy for the lower tolerance values is significantly less as compared to the other measures of interest. 



\subsubsection{Half Power Beam Width}

For this measure, we consider the tolerance values of 1º, 0.5º and 0.25º. From Table III we observe that the MLPNN has corresponding accuracy of 99.5\%, 97.3\% and 79.2\%, respectively. Further, the CNN has accuracy measures of 98.8\%, 92.6\% and 61.8\%, respectively. Note that, the trend for the accuracy values is similar to that observed for the other measures of interest (Sections V.B.1-V.B.3).
\\
\newline
\begin{figure}[t]
    \centering
    \includegraphics[width=0.48\textwidth]{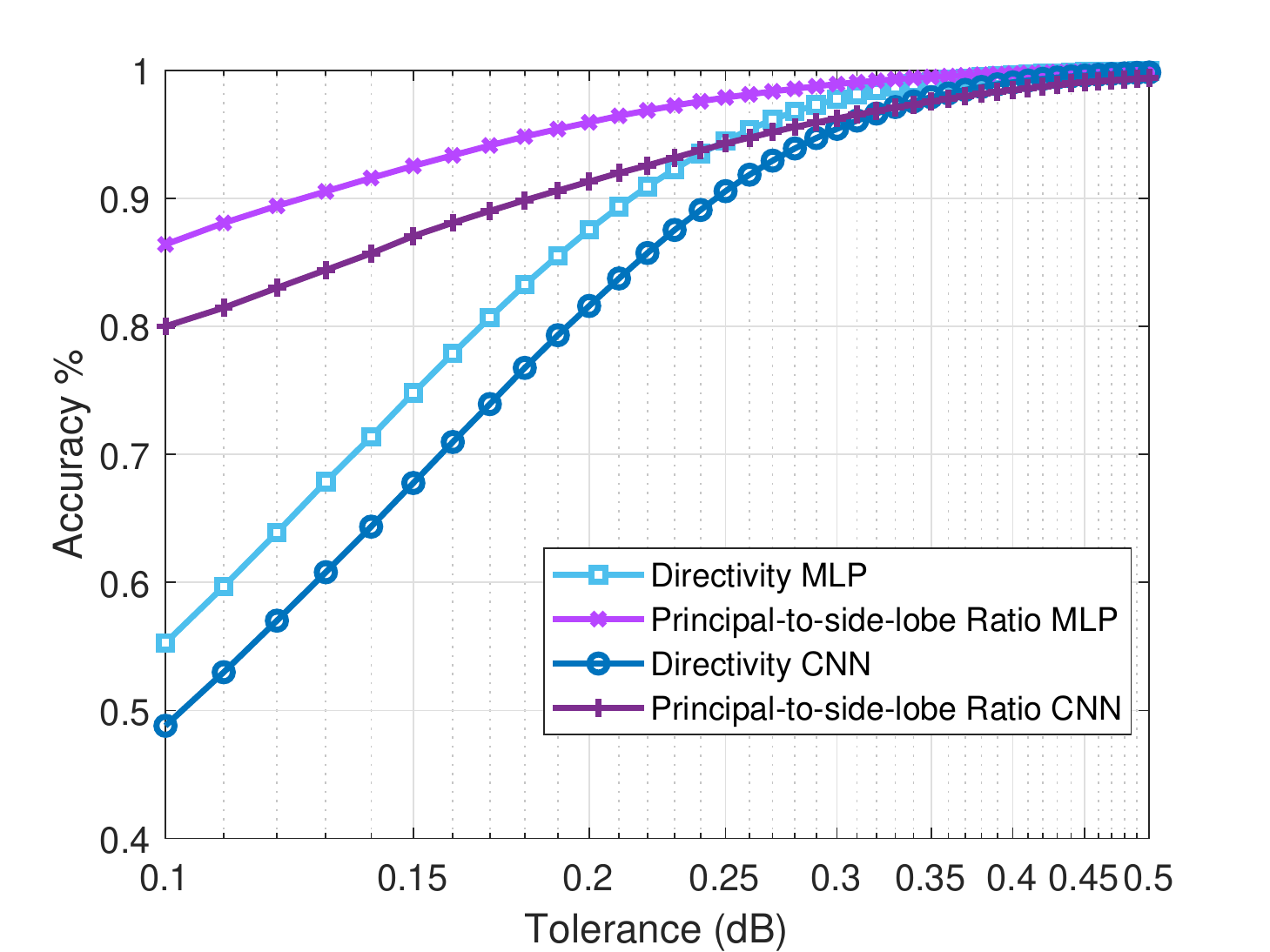}
    \caption{Accuracy vs tolerance in dB for both MLPNN and CNN. The curves shown correspond to Directivity and Principal-to-side-lobe Ratio.}
    \label{fig:accvstoldb}
\end{figure}

\begin{figure}[t]
    \centering
    \includegraphics[width=0.48\textwidth]{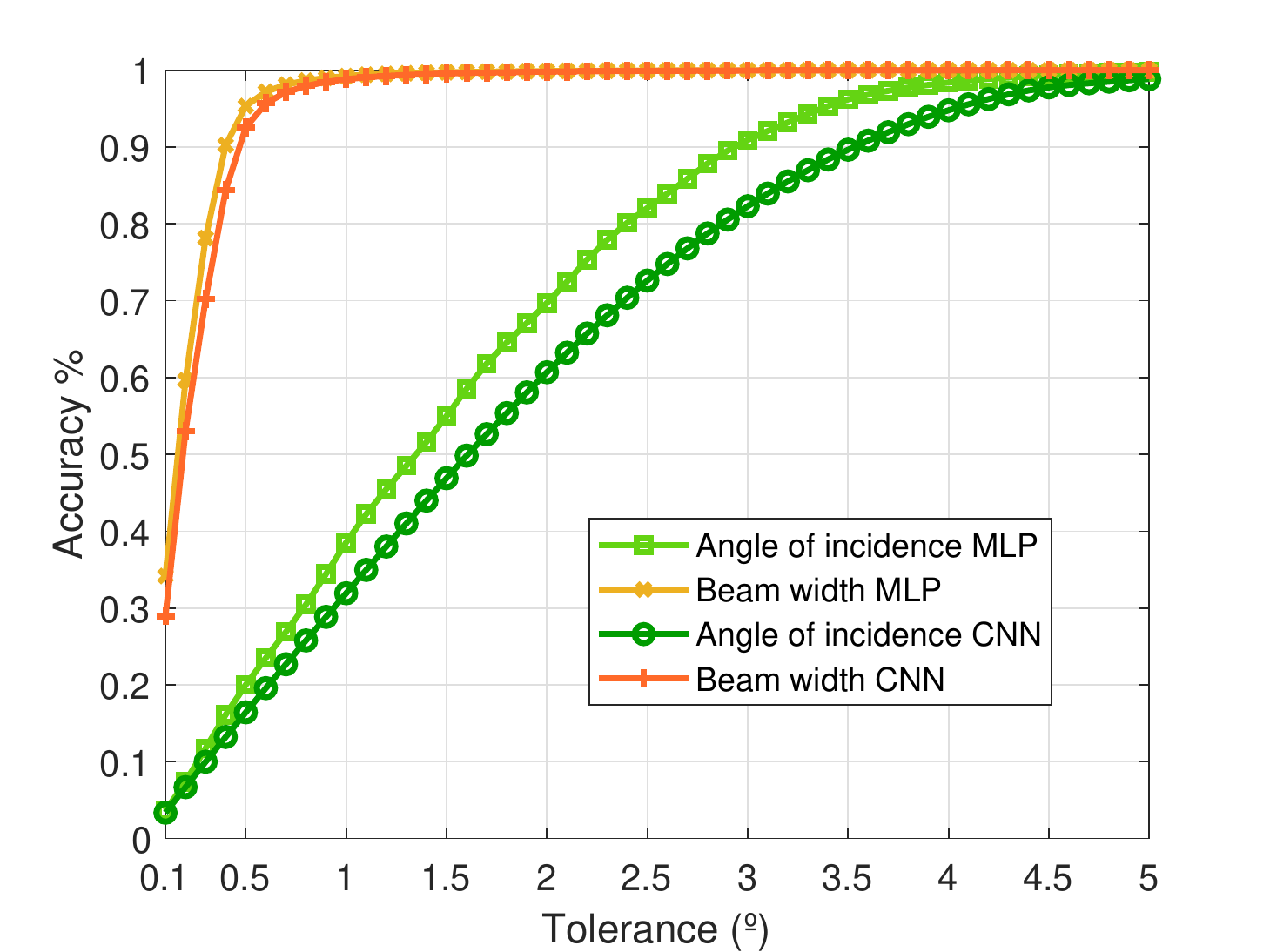}
    \caption{Accuracy vs tolerance in degrees for both MLPNN and CNN. The curves shown correspond to Angle of maximum radiation and Beam width.}
    \label{fig:accvstoldeg}
\end{figure}
\noindent And so from the discussions so far, we can deduce that the proposed methodology is able to accurately predict the reflected beam radiation pattern or the measures that can fully characterize the same. To further reinforce this idea, we present Figs. \ref{fig:accvstoldb} and \ref{fig:accvstoldeg}. Concretely, Fig. \ref{fig:accvstoldb} shows in detail the evolution of the accuracy of the predictions for the \textit{Directivity} and \textit{Principal-to-side-lobe ratio} as the tolerance in dB grows, for both MLP and CNN cases. We observe that the trend for the accuracy is exactly what we have deduced through our discussions in sections V.B.1-V.B.4. Further, we see that as the tolerance approaches 0.5 dB, the accuracy of CNN predictor approaches that of the MLP.  

Lastly, Fig. \ref{fig:accvstoldeg} illustrates the evolution of the accuracy of the predictions for the \textit{Angle of incidence} and \textit{Half Power Beam Width} as the tolerance in degrees grows, for both MLP and CNN cases. Again, here we observe that the accuracy percentage improves as the tolerance is increased. However, we also notice that the beam width prediction approaches near 100\% accuracy at very low tolerance values, whilst the \textit{angle of radiation} measure necessitates higher tolerance limits for the predictors to achieve better accuracy.

\section{Conclusion}
\label{sec:conclusion}

In this paper, we have presented a novel data driven methodology, wherein we utilize a NN based approach for characterizing the reflected beam radiation pattern from a metasurface. One of most important advantages of such an approach is that, while its accuracy is close to that of the full wave simulator approaches, the time complexity to achieve the same is significantly smaller. In addition, it can also serve as a methodology that enables self healing characteristics and facilitates maintenance aspects of MSFs in the 6G wireless network environment. 

And so, as part of this methodology, we have provisioned an incremental design framework. Through this framework we analyzed three specific scenarios, wherein we develop an estimation of the full radiation pattern for a non-tunable MSF, a globally tunable MSF and then scaling it to a locally tunable MSF. Further, through our analysis we have demonstrated the efficacy of the NN based approaches. Concretely, it was observed that the NN based approaches could predict the radiation pattern with a very high accuracy in a significantly reduced time frame as compared to the full wave simulator counterparts. 

Moreover, through the third scenario, we demonstrated that our novel CNN based prediction framework performs as well as the fully connected MLPNN framework. However, it does so with a significantly reduced computational complexity as compared to the fully connected MLPNN framework. This will especially be critical, when the framework is scaled up to even larger MSF configurations. 

Lastly, in this work, through the last scenario, we have also provisioned a first study, wherein, instead of estimating the entire radiation pattern, we have predicted the most important parameters that govern any radiation pattern, i.e., Directivity, Principle-to-side lobe ratio, Angle of maximum radiation and Beam width. This process will not only ensure the required reliability in estimation but it will also allow for a faster convergence time for such estimations.  

\section*{Appendix A}

In the the MLP case, the typical regularization methods utilized are L1 and L2. These methods add a new term in the cost function that sums all the non-zero weights. The significance of this term is governed by the regularization parameter $\lambda$. And while, the main advantage of L1 regularization is that it forces sparsity into the models by forcing most of the weights to zero, for our case we do not require feature selection since the number of input pixels is quite limited. Consequently, L2 regularization is selected and the new loss function is defined as:
\begin{equation*}
L(X,y) = MSE(X,y) + \lambda \sum_{i=1}^{N}w_i^{2}
\end{equation*}
where $X$ represent the input images, $y$ the target metrics, $\lambda$ the regularization parameter, N the total number of neurons of the MLP and $w$ each weight of all the layers.

Furthermore, for selecting the optimal L2 regularization parameter $\lambda$, we performed a 10-fold cross-validation. For each parameter value, we split the dataset into 10 groups, and for each group we train a model with the remaining 9 groups, and validate with the selected group. Then, the validation errors of each combination are averaged. Finally, the parameter value that provides the lowest cross-validation Mean Squared Error (MSE) is selected for the final model.

\bibliographystyle{IEEEtran}
\bibliography{IEEEabrv,references,new-refs}
%


\end{document}